\documentclass[12pt]{article}

\usepackage{amsfonts}
\usepackage{amssymb}
\usepackage{amsthm}
\usepackage{upref}

\newcommand{\sss}{\setcounter{equation}{0}}
\newtheorem{theorem}{THEOREM}[section]

\newtheorem{remark}[theorem]{REMARK}

\newtheorem{prop}[theorem]{PROPOSITION}

\newfont{\BBFONT}{msbm10 scaled 1200}
\newcommand{\ere}{ {\mathbb R}}

\newcommand{\ese}{{\mathbb S}}

\newcommand{\sym}{a(y,\omega;\lambda)}
\def\beq{\begin{equation}}
\def\ene{\end{equation}}

\def\qed{\ifhmode\unskip\nobreak\fi\ifmmode\ifinner
\else\hskip5pt\fi\fi\hbox{\hskip5pt\vrule width4pt height6pt
depth1.5pt\hskip1pt}}
\def\scat{s(\nu,\omega; \lambda)}
\def\sc0{s_0(\omega,\nu; \lambda)}

\begin{document}
\baselineskip=20 pt
\parskip 6 pt

\title{ Inverse Scattering at a Fixed Energy for Long-Range   Potentials
 \thanks{ Research partially supported by CONACYT
under Project P42553­F and by the European Group of Research
SPECT.}}
 \author{ Ricardo Weder\thanks{ †Fellow Sistema Nacional de Investigadores.}
\\Instituto de Investigaciones en Matem\'aticas Aplicadas y en Sistemas \\Universidad Nacional Aut\'onoma de
M\'exico \\
Apartado Postal 20-726, M\'exico DF 01000 \\weder@servidor.unam.mx
\and Dimitri Yafaev \\IRMAR, Universit\'e de Rennes I\\
 Campus de Beaulieu, 35042 Rennes, Cedex, France
\\ yafaev@univ-rennes1.fr}
\date{}
\maketitle
\begin{center}
\begin{minipage}{5.75in}
\centerline{{\bf Abstract}}
\bigskip

In this paper we consider the  inverse scattering problem at a
fixed energy for the Schr\"odinger equation with  a long-range
potential  in $\ere^d, d\geq 3$.  We prove that
the long-range part can be   uniquely reconstructed from the leading
forward singularity  of the scattering amplitude at some
positive energy.

\end{minipage}
\end{center}

\section{Introduction}\sss

  Our goal is to study the
inverse scattering problem at a fixed energy for the Schr\"odinger
equation with a long-range potential. We give a method for the
unique reconstruction of the long-range part of the potential at   infinity.
The reconstruction of the short-range part is similar to the procedure suggested in \cite{wy} where
the same problem for short-range potentials was treated. To a certain extent,
this paper can be considered as   continuation
 of \cite{wy}, but we do not dwell here upon reconstruction of the short-range part of a long-range potential.

For a short survey of different formulations of the inverse
scattering problem see \cite{wy}. Here we just mention the
contributions to long-range inverse scattering. Isozaki and Kitada
\cite{ik} consider potentials that satisfy
\beq
 \left| \partial^\alpha V(x)\right| \leq C_\alpha   (1+|x|)^{-\rho-|\alpha|}, \quad  \rho > 1/2, \quad
 \hbox{for all $\alpha$}.
 \label{1.1}
 \ene
 Using stationary
methods, they  proved that the high-energy limit $\lambda\rightarrow\infty$ of the scattering
matrix  $S(\lambda)$ determines uniquely a  potential and give  a method for its  reconstruction.
A similar result was obtained by Yafaev \cite{yaf4} for an arbitrary $ \rho > 0$.
More precisely,   the Fourier transform of the
potential is reconstructed  in \cite{ik}, \cite{yaf4} by taking the  high-energy limit of the scattering amplitude with
fixed momentum transfer.  Enss and Weder \cite{ew} proved a similar high-energy
uniqueness result and gave a reconstruction method using a time-dependent
approach. They consider a large class of long-range potentials that
are allowed to have singularities as well as N-body potentials. In \cite{ew} the $X$-ray transforms  of the potentials are uniquely
reconstructed from the action of the scattering operator on appropriate
high-energy states.

As mentioned above, we are interested in this paper in the inverse
scattering problem at a fixed energy in the case of long-range
potentials in $\ere^d$, $d \geq 3$. Surprisingly, this  problem has received little
attention. In \cite{jo} it was proven that the scattering matrix
at a fixed energy uniquely determines the asymptotics of a
short-range potential on the background of a Coulomb potential,
what actually is a different problem. We also mention
paper \cite{ni}, where using the method of \cite{ew} the
asymptotics of a long-range potential is uniquely reconstructed
if the scattering matrix is known on some (perhaps arbitrarily
small) energy interval.

We suppose that $V\in C^\infty(\ere^d)$, $d \geq 3$, and that, for sufficiently large $|x|$,
\beq
V(x)= \sum_{j=1}^N V_{j}(x)+ V_{sr}(x),
\label{4.1AS}
\ene
where $V_j \in C^\infty(\ere^d \setminus\{0\})$ is a homogeneous function of order $-\rho_j$, i.e.,
$V_j(tx)= t^{-\rho_j} V_j(x)$ for all $t>0$, with $ 1/2 < \rho_1 < \rho_2 < \cdots < \rho_N \leq 1$, and $V_{sr} \in
C^\infty(\ere^d)$ is a short-range potential that satisfies (\ref{1.1}) for some  $\rho_{sr} >1$.
Our objective   is to uniquely reconstruct all functions $V_j(x)$, $j=1,\ldots, N$, if only the leading forward singularity of the
scattering amplitude    is known at some fixed  energy $\lambda>0$. Of course, we do not aim
to reconstruct the whole potential, as is the case when the
high-energy limit of the scattering matrix  is known. The key
issue here is that the forward singularity of the scattering
 amplitude contains all the information about the behaviour of the
potential at infinity, and this allows us to uniquely reconstruct
its asymptotic expansion at infinity.
In particular, the leading singularity is sufficient for reconstruction of all long-range terms. Technically, we follow sufficiently closely our previous paper \cite{wy}, where
short-range potentials that satisfy (\ref{1.1}) with $ \rho > 1$ were considered.

 Under condition (\ref{1.1}) the scattering matrix $S(\lambda) $, where $\lambda>0$ is the energy
 of a quantum particle, is a unitary operator on $L^2(\ese^{d-1})$. Formally, the scattering matrix can be considered as an integral operator, that is
\[
 \left(S(\lambda)f \right)(\nu)= \int_{\ese^{d-1}}\,
 \scat\,f(\omega)\, d \omega,
 \]
with integral kernel (the scattering amplitude) $\scat$. Here $\omega$ is the direction of the incident beam of particles and $\nu$ is the direction of observation. We emphasize that our definition of the scattering amplitude is somewhat different from the short-range case where the integral kernel of $S(\lambda)$ and the scattering amplitude differ by the Dirac delta-function. In the long-range case the delta-function disappears from  the integral kernel. As is well known \cite{ag}, the scattering amplitude is $C^\infty$-function away from the diagonal $ \nu=\omega$,
but its diagonal singularity is very wild \cite{yaf3, yaf4}.

Actually, it is more convenient (especially, in the long-range case) to consider  the scattering matrix  as a pseudodifferential
 operator. It means that
  \beq
 \left(S(\lambda) f \right)(\nu)= (2\pi)^{-d+1} k^{ d-1 }\,\int_{\Pi_\omega}\,
 \int_{\ese^{d-1}}\,
e^{- i  k\langle y,\nu\rangle}
\, a(y,\omega;\lambda)\,
f (\omega)\, dy  d \omega, \quad k=\lambda^{1/2},
\label{1.2X}
\ene
where $\Pi_\omega$ is the hyperplane in $\ere^d$ orthogonal to $\omega$,  $y\in \Pi_\omega$ is known in the physics literature as the impact parameter  and $a(y,\omega;\lambda)$ is the right symbol  of the pseudodifferential operator $S(\lambda)$.  It is related
to  the scattering amplitude by the formula
\beq
\scat= (2\pi)^{-d+1} k^{ d-1 }\,\int_{\Pi_\omega}\,
e^{- i  k\langle y,\nu\rangle}
\, a(y,\omega;\lambda)\, dy.
\label{1.3}\ene
Here and below we use freely the terminology of  pseudodifferential calculus (see, e.g., \cite{Sh}, \cite{tr1}).  For example, expressions such as (\ref{1.2X}) or (\ref{1.3}) are understood as oscillating integrals.
 Note that     our  definitions differ from the standard ones
by the factor $-k$ in the phase in (\ref{1.2X}) or (\ref{1.3}).  The fact that the scattering matrix is well defined as a pseudodifferential
 operator was proven in \cite{yaf4}. In particular, it was shown there that
its symbol $a(y,\omega;\lambda)$ belongs to the H\"{o}rmander class $\mathcal S^0_{\rho,1-\rho}=
 \mathcal S^0_{\rho,1-\rho}(T^* \ese^{d-1})$ if $\rho<1$ and to the to class $\mathcal S^0_{1-\varepsilon, \varepsilon}$ for any $\varepsilon>0$  if $\rho =1$.
 Here $T^* \ese^{d-1}$ is the cotangent bundle of the unit sphere, that is the set of points $(y,\omega)$ such that $\omega \in \ese^{d-1}$ and $y\in\Pi_\omega$. The class $S^0_{\rho,1-\rho}$ fits in  the standard pseudodifferential calculus exactly in the case $\rho>1/2$.

 We proceed from the results of \cite{yaf3, yaf4} where it was shown that the principal symbol $a_{0}(y,\omega;\lambda)$ of $S(\lambda)$ is given by the equation
 \beq
a_{0}(y,\omega;\lambda)= e^{-i (2k)^{-1} \Phi(y,\omega )}
\label{1.6V}
\ene
where
\beq
\Phi(y,\omega)=\Phi(y,\omega;V)=   \int_{-\infty}^\infty (V(y+t\omega)-V(t\omega))\,dt.
\label{1.7}\ene

Roughly speaking, our approach consists of the following steps.

1. Given $S(\lambda)$, we find its principal symbol $a_{0} (y,\omega;\lambda)$. Actually, it suffices for us to know the operators
 \beq
S_{\omega_{0}}(\lambda) = \varphi_{\omega_{0}}S(\lambda) \varphi_{\omega_{0}},
\label{1.6R}\ene
where $\omega_{0}\in {\Bbb S}^{d-1}$ is an arbitrary point and $\varphi_{\omega_{0}}$ is multiplication by  the function $\varphi_{\omega_{0}}\in C^\infty ({\Bbb S}^{d-1})$ such that $\varphi_{\omega_{0}}(\omega)=1$ in some,  arbitrary small, neighborhood $O_{\omega_{0}}$ of the point $\omega_{0}$. Then we use the fact that  $a_{0} (y,\omega;\lambda)$ coincides with the principal symbol of the pseudodifferential operator $S_{\omega_{0}}(\lambda)$ for $\omega\in O_{\omega_{0}}$.

2. For the  reconstruction of the long-range part of $V$, it suffices to know function (\ref{1.7}). Clearly, if $V$ is asymptotically homogeneous of order $-\rho$, then $\Phi$ is an asymptotically   homogeneous function of order $-\rho +1$ of the variable $y$
(except the case $\rho=1$ when $\Phi $ has  a logarithmic behavior at infinity).
Under assumption (\ref{4.1AS}) the  contributions to $\Phi$
  of different functions $V_{j}$
    can clearly be separated in (\ref{1.7}).
 Then we can directly reconstruct   $V_{j}$ by the inversion of the Radon transform (in some two-dimensional plane not passing through the origin).

Eventually, our method extends to potentials that satisfy (\ref{1.1}) with $\rho >0$. However, in the general
case two new additional difficulties should be taken into account. The first is that
the phase function in (\ref{1.6V}) is given by a more complicated formula than (\ref{1.7}) although (\ref{1.7}) remains its first approximation (see \cite{Roux}, \cite{yaf4}).   The second difficulty is that for $\rho \leq 1/2$ the symbol of the scattering matrix is oscillating too rapidly so that
the standard pseudodifferential operator calculus cannot be applied. In this case one has to use more specific results for pseudodifferential operators with oscillating
symbols \cite{yaf1}.

 The paper is organized as follows. In Section 2, we recall different definitions of the wave operators in the long-range case.
The scattering operator and the scattering matrix are also introduced there.
Following \cite{yaf3, yaf4},    we give in Section 3 the  description of leading forward singularity of the   scattering amplitude.  The
classical inversion formula for the Radon transform is   recalled in Section 4.
In Section 5 we   uniquely reconstruct  the long-range part of the potential.

\section{ Long-Range Scattering}\sss

Here we recall some basic definitions
of long-range scattering theory;  see, e.g., \cite{yaf2}, for more details.
 We consider the Schr\"odinger operator
\[
H=-\Delta+V(x)
\]
with potential $V(x)$ in the space $L^2(\ere^d)$
 where $d\geq 2$. If $V$ is a real and bounded function, then the Hamiltonian $H$ is well defined on the Sobolev class $\mathsf{H}^2(\ere^d)$ and is
self-adjoint   in the space $L^2(\ere^d)$. Let us denote by $H_0=-\Delta$ the ``free" Hamiltonian
corresponding to the case $V=0$.  Under   assumption
(\ref{1.1})  the operator $H$ has no singular  continuous spectrum,        its absolutely continuous spectrum coincides with $[0,\infty)$, and
its negative spectrum consists of eigenvalues.

Since the usual wave operators do not exist for $\rho\leq 1$,   the large-time asymptotics of $e^{-itH}\, u$ for vectors $u$ from the absolutely continuous subspace  of $H$ is described in terms of the modified free evolution. There are several possibilities to construct it. For example, in coordinate representation the modified free evolution $U_0(t)$ is defined in  \cite{Y2}  by the equation
\beq
(U_0(t)u )(x)= e^{i\Xi(x,t)}\, (2it)^{-d/2} \hat{u}(x/(2t)),
\label{2.2}
\ene
where
\beq
\Xi(x,t) = (4t)^{-1}\,|x|^2 -t\int_0^1 V(sx)\, ds,
\label{2.3}
\ene
and
$$
\hat{u}(\xi)=  (F u)(\xi) = (2\pi)^{-d/2}\, \int_{\ere^d}\, e^{-i\langle x,\xi\rangle}\,
u (x)\, dx
$$
is the Fourier transform of $u$. Then  the modified wave operators
\beq
W_{\pm} = \hbox{s}-\lim_{t \rightarrow \pm \infty} e^{ itH} U_0(t)
\label{2.4}
\ene
exist and have the intertwining property $H W_{\pm}= W_{\pm} H_0$. Moreover, they are asymptotically complete,
i.e., their ranges coincide with  the absolutely continuous subspace  of $H$.

Equivalently,  the modified free dynamics  can   be defined (see \cite{Do} and \cite{BM}) in momentum representation by the equation
\[
(F \tilde{U}_0(t) u)(\xi)= e^{- i | \xi | ^2 t}\, e^{- i \int_0^t V(2\xi s)\, ds} \hat{u}(\xi).
\]
Although the operators $U_0(t)$ and $\tilde{U}_0(t)$ do not coincide,   $W_{\pm}$ equals the wave operator
\[
\tilde{W}_{\pm} = \hbox{s}-\lim_{t \rightarrow \pm \infty} e^{ itH} \tilde{U}_0(t).
\]
Still another possibility is to define
  the modified free dynamics   by the introduction of an appropriate time-independent modifier \cite{IK3,ik}. In this case the role of $U_{0}(t)$ in the definition of $W_{\pm}$ is played by the operator $J_{\pm} e^{ -itH_{0}}$ where $J_{\pm}$ is a specially constructed pseudodifferential operator.

 Given the wave operators,  the scattering operator and the scattering matrix are defined exactly as in the short-range case. It follows from properties of the wave operators that the scattering operator
\beq
 {\mathbf S} = W^{\ast}_+ \, W_-
 \label{2.5}
\ene
commutes with $H_{0}$ and   is unitary in the space $L^2 (\ere^{d})$.
Let $\ese^{d-1}$ be the unit sphere in $\ere^d $,
  $\ere^+  = (0,\infty)$ and let $L^2\left(\ere^+, L^2(\ese^{d-1})\right)$
  be the $L^2$-space of  functions defined on $\ere^+$ with values   in $L^2(\ese^{d-1})$. Define the unitary
  operator
$$
{\cal F}: L^2 (\ere^{d})
\rightarrow
L^2\left(\ere^+, L^2(\ese^{d-1})\right)
$$
by the equation
$$
({\cal F}u )(\omega;\lambda)=2^{-1/2} \lambda^{(d-2)/4} \hat{
u}(\lambda^{1/2} \omega).
$$
 The spectral parameter $\lambda$   plays the role of the
 energy of a quantum particle. Then $({\cal F}H_{0}u )( \lambda)=\lambda ({\cal F} u )( \lambda)$ and
$$
({\cal F}{\mathbf S} u )( \lambda)=S(\lambda)  ({\cal F} u )( \lambda).
$$
The unitary operator $S(\lambda): L^2(\ese^{d-1})\rightarrow
L^2(\ese^{d-1})$ is known as the scattering matrix at energy
$\lambda$.

Of course, the definition of the modified free dynamics and hence of modified wave operators is not unique. However, the freedom in their choice is rather limited. For example, in definition (\ref{2.2}) one can add to $\Xi (x,t)$ an arbitrary (smooth) function which behaves as $\theta_{\pm}(x/(2t))$ for $t\rightarrow\pm\infty$. Then the wave operator (\ref{2.4}) is replaced by $W_{\pm}e^{i\theta_{\pm}({\mathbf p})}$
where ${\mathbf p}=-i \nabla$ and $e^{i \theta_{\pm}({\mathbf p})}=F^\ast e^{i\theta_{\pm}(\xi)}F$
is    multiplication by $e^{i\theta_{\pm}(\xi)}$ in the momentum representation.
 It follows that the scattering operator (\ref{2.5}) is replaced by
$
\tilde{\mathbf S} =  e^{-i\theta_+({\mathbf p})}\, {\mathbf S} \, e^{i\theta_-({\mathbf{p}})}
$
and the scattering matrix   $S(\lambda)$ is replaced by
\beq
\tilde{S} (\lambda)=  e^{-i \theta_+ (\sqrt{\lambda}\,\,\cdot)}\, S(\lambda)\, e^{i\theta_-(\sqrt{\lambda}\,\,\cdot)}.
\label{2.12}
\ene
 In particular, if $V$ is a sum of long-range $V_{lr}(x)$ and short-range $V_{sr}(x)$ functions, then  $V$ can be replaced by  $V_{lr}(x)$ in (\ref{2.3}). In this case
 \[
 \theta_{\pm}(\xi)=2^{-1} \int_{0}^{\pm\infty}V_{sr}(\xi s)ds.
 \]

We accept below that the scattering matrix is defined in terms of wave operators (\ref{2.4}) with phase function (\ref{2.3}).

 \section{ The Structure of the Scattering Matrix}
\sss

\noindent   We need to know only the leading singularity of the scattering amplitude. The following result was
essentially obtained in \cite{yaf3}, but it is also a consequence of more general results of \cite{yaf4} where a
complete description of all singularities was found.

\begin{theorem}\label{th5.1P}
Suppose that estimate $(\ref{1.1})$ holds for all $\alpha$. Then the scattering matrix $S(\lambda) $ is a pseudodifferential operator on the unit sphere $\ese^{d-1}$
with the symbol
\beq
\sym= e^{-i (2k)^{-1} \Phi(y,\omega )}\,(1+b(y,\omega;\lambda)),\quad k=\lambda^{1/2},
\label{1.6}
\ene
where $\Phi  $ is function $(\ref{1.7})$
and $b\in \mathcal S^{-2\rho +1}_{1,0}$ if  $\rho<1$ and $b\in \mathcal S^{-1+\varepsilon}_{1,0}$ for any $\varepsilon>0$ if  $\rho=1$.
\end{theorem}

\begin{remark}
\label{reM}
{\rm
 Scattering matrix  (\ref{2.12}) is also a pseudodifferential operator with the symbol
\beq
\tilde{a}(y,\omega;\lambda) = e^{-i (2k)^{-1}\tilde{\Phi}(y,\omega )  }\,(1+\tilde{b}(y,\omega;\lambda)),
\label{1.6A}
\ene
where
\beq
\tilde{\Phi}(y,\omega )=\Phi(y,\omega ) +2k \theta_{+}(k\omega) - 2k\theta_{-}(k\omega)
\label{1.6B}
\ene
 and $\tilde{b}  $ belongs to the same class as the function $b$ in Theorem~\ref{th5.1P}.
 }
\end{remark}

\bigskip

\section{\bf   Inversion of the Radon transform}
\sss

\noindent To solve our   inverse scattering problem we use  the two-dimensional Radon
transform (see, e.g., \cite{hel}).
 Here we  briefly recall some of its properties. For $v\in {\cal S}^{-\rho}( \ere^2)$, $\rho>1$, the Radon
 transform, or $X$-ray transform, which is the same in  two dimensions,  is defined by the   formula
\[
r(y,\omega;v)= \int_{-\infty}^\infty v(\omega t+y) dt,\quad \omega\in \ese, \quad \langle \omega, y\rangle =0.
\]
It is clear that, $r(y, \omega)=r(y, -\omega)$. The Fourier transform $\hat{v}$  of $v$ and hence the function $v$ itself
 can be reconstructed from its Radon transform in the following way. Let $\omega_{\xi}$ be one of the two unit
 vectors such that $ \langle \omega_{\xi} , \xi\rangle =0$. Hence,
\beq
\hat{v}(\xi)=(2\pi)^{-1}\int_{-\infty}^\infty e^{- i |\xi | s} r( s\hat{\xi}, \omega_{\xi} ;v) ds,
\quad \hat{\xi}=\xi |\xi |^{-1}.
\label{3.4}
\ene

Below we apply this method for the reconstruction of   a homogeneous
function $V \in C^\infty (\ere^d\setminus \{0\})$
 of order $-\rho< 0$ from the integral $ \Phi(y,\omega )$ defined by formula (\ref{1.7}). We suppose that it is known for all $\omega\in \ese^{d-1}$ and all $y\in \Pi_{\omega}$,    $y\neq 0$. Actually, it suffices to know the function $\nabla\Phi(y,\omega )$ where $\nabla=\nabla_{y}$. For  an arbitrary $x\in\ere^d \setminus \{0\}$, we shall find $V(x)$. Let us  fix the coordinate system in such a way that
the first axis is directed along $x$, i.e., $x=(x_1,0,\cdots , 0)$, and consider some two-dimensional plane $\Lambda_{x}$ orthogonal to $x$. Suppose that $\omega\in\Lambda_{x}$, $|\omega|=1$. Differentiating (\ref{1.7}) with respect to $y_{1}$, we find that
$$
\partial_{y_{1}} \Phi (y,\omega;V)=  \int_{-\infty}^\infty \,
 \partial_{y_{1}} V (y+t\omega)\, dt.
$$
 For   $\overline{y}\in\Lambda_{x}$ such that $\langle \overline{y},\omega\rangle =0$,  set
$$
v_{x}(\overline{y}):= \partial_{x_{1}}  V (x+\overline{y}).
$$
Then, for   all $\omega\in\Lambda_{x}$, $|\omega|=1$,  and all $\overline{y}\in\Lambda_{x}$,
 $\langle \overline{y}, \omega \rangle =0$,
\[
r(\overline{y}, \omega ;v_{x})=  \partial_{y_{1}} \Phi ( x+\overline{y},\omega;V) .
\]
Since $x+\overline{y}
\neq 0$, the function $v_{x} \in {\cal S}^{-\rho-1}({\Lambda_{x}})$ so
that we can recover  $v_{x}$
 and, in particular, $v_{x}(0)=  \partial_{x_{1}} V  (x)$ by formula  (\ref{3.4}).
 Then, integrating we reconstruct
$$
V(x) = -\int_{x_1}^\infty \partial_{s}V (s,0,\cdots ,0)\, ds.
$$
 In particular, we have proven the following proposition.

  \begin{prop}
Let $V\in C^\infty (\ere^d\setminus \{0\}), d \geq 3,$ be a homogeneous function of
order  $-\rho< 0$. If $\nabla\Phi ( y,\omega )=0$ for all $\omega\in\ese^{d-1}$ and all $y\in \Pi_{\omega}$, $y
\neq 0$, then  $V(x)=0$.
\end{prop}

\bigskip

\section{Reconstruction theorem}
\sss

Reconstruction of the long-range part  of the potential requires only the knowledge  of the leading term in the asymptotics of the symbol $a(y,\omega;\lambda)$ as $|y|\rightarrow\infty$. The necessary result is formulated in Theorem~\ref{th5.1P}.

First,  we reconstruct the symbol $\sym$ from a family of the operators $S_{\omega_{0}}(\lambda)$ defined by equation (\ref{1.6R}). Here we use the fact that  $a  (y,\omega;\lambda)$ coincides with the   symbol of the pseudodifferential operator
$ S_{\omega_{0}}(\lambda) $   for $\omega\in O_{\omega_{0}}$.

Then, by equation (\ref{1.6}),
\beq
2k a(y,\omega;\lambda) ^{-1} \nabla a(y,\omega;\lambda) = -i \nabla\Phi (y,\omega )+
2k (1+b(y,\omega;\lambda))^{-1} \nabla b(y,\omega;\lambda)   .
\label{4.1}
\ene
Under  condition (\ref{4.1AS}),  $\nabla b  (1+b)^{-1} \in \mathcal S^{-p} $
where $p=2\rho_{1}> 1$ if $\rho_{1}<1$ and $p $ is any number smaller than $2$ if $\rho_{1}=1$,
whereas
\beq
\nabla \Phi (y,\omega; V) =  \sum_{j=1}^N \nabla \Phi (y,\omega; V_{j}) +
\nabla \Phi (y,\omega; V_{sr}) .
\label{4.2}
\ene
Here  $ \nabla \Phi (y,\omega; V_{j})$, $ j=1,\cdots,N$, are homogeneous functions of orders $-\rho_j \geq -1$  and $\nabla \Phi (\cdot ,\omega; V_{sr})
\in \mathcal S^{-\rho_{sr}}$ with $ -\rho_{sr}< -1$. Thus, given $a(y,\omega;\lambda)$, we single out in expression (\ref{4.1}) all homogeneous terms of orders $\geq -1$. This yields us the functions  $\nabla \Phi (y,\omega; V_{j})$. Finally, each one of the $V_j$, $j=1,\cdots,N$,
is uniquely reconstructed from $\nabla \Phi  (y,\omega; V_j)$, as explained in the previous  section.

Note that the number, $N$, of long-range terms, as well as their order of homogeneity, $-\rho_j$, $j=1,\cdots,N$,
are obtained in the reconstruction process.We do not need to know them {\it a priori}.

In this way we have proven our main result.

\begin{theorem}\label{th4.1}
Suppose that $V\in C^\infty(\ere^d)$, $d \geq 3$, and that for sufficiently large $|x|$ equality $(\ref{4.1AS})$ is true.
Here $V_j \in C^\infty(\ere^d \setminus\{0\})$ is a homogeneous function of order $-\rho_j$ with $ 1/2 < \rho_1 < \rho_2 < \cdots < \rho_N \leq 1$, and $V_{sr} \in
C^\infty(\ere^d)$ is a short-range potential that satisfies $(\ref{1.1})$ for some  $\rho_{sr} >1$.
Then, for an arbitrary $\lambda >0$,  any family of the operators $S_{\omega_{0}}(\lambda)$  uniquely determines each $V_j$, $j=1,\cdots, N$. Moreover, the functions $V_j$ can be reconstructed
from formulae $(\ref{4.1})$, $(\ref{4.2})$ by the inversion of the $X$-ray transform.
\end{theorem}

\begin{remark}
\label{re6.4}
{\rm Let $\tilde{S}(\lambda)$ be the scattering matrix defined by formula (\ref{2.12}).
Considered as a pseudodifferential operator   it has symbol (\ref{1.6A}). According to (\ref{1.6B})
\[
\nabla\tilde{\Phi} (y,\omega )= \nabla \Phi (y,\omega)
\]
so that in view of (\ref{4.2}) the functions $V_{j}$, $j=1,\ldots, N$,  can be reconstructed from the function $\tilde{\Phi}$.
Hence, we can reconstruct the long-range part of the potential from any one of the possible choices
of scattering matrices, that correspond to different modified free dynamics.
}
\end{remark}

\end{document}